\documentclass[12pt]{article}

\usepackage{amssymb}
\usepackage{amsmath}

\usepackage{epsfig}
\begin{document}

\title{\bf Hadron multiplicity in $\mathbf{e^+e^-}$ events induced by
top quark pairs at the ILC energy}

\author{A.V. Kisselev\thanks{E-mail: alexandre.kisselev@ihep.ru} \
and V.A. Petrov\thanks{E-mail: vladimir.petrov@ihep.ru} \\
\small Institute for High Energy Physics, 142281 Protvino, Russia}

\date{}

\maketitle

\thispagestyle{empty}

\bigskip

\begin{abstract}
The average charged hadron multiplicity in the $e^+e^-$ events
with the primary $t\bar{t}$-pair at the collision energy 500 GeV,
as well as the average multiplicity of charged hadrons from the
top quark are calculated in QCD to be  $86.7 \pm 1.11$ and $41.0
\pm 0.54$, respectively.
\end{abstract}


\section{Introduction}

Experiments at LEP and SLAC revealed, besides other important
results, quite interesting feature of the hadron multiple
production dependent on the mass of the ``primary'' (anti)quarks
which launch the process of the QCD evolution. It appeared that
differences between the light and heavy quark-induced
multiplicities become energy-independent. QCD calculations
describe the phenomenon quite well.

Certainly, LEP could not give the information on the events
induced by the top quarks. Recent discussions of the ILC project
give us occasion to provide QCD predictions concerning the hadron
multiple production in the events with primary $t$-quarks.

We manage to calculate the average hadron multiplicity
$N_{t\bar{t}}$ in the $e^+e^-$ events with $t\bar{t}$ pair at the
collision energy of the ILC with the following prediction:
\begin{equation}\label{N_tt_prediction}
N_{t\bar{t}} (W = 500 \mathrm{\ GeV}) = 86.67 \pm 0.55 \;.
\end{equation}
We also theoretically calculated the average hadronic multiplicity
from the top quark:
\begin{equation}\label{n_t_prediction}
n_t  = 41.03 \pm 0.27 \;.
\end{equation}
Both values correspond to the average value of the top mass $m_t =
170.9$.  Everywhere below, it is assumed that we deal with
\emph{average} multiplicities of \emph{charged} hadrons.

The paper is organized as follows. In order to make our
calculations of the hadron multiplicity in top quark events more
easy for understanding, we consider first the multiple hadron
production in $c\bar{c}$ ($b\bar{b}$) events. The hadron
multiplicities in $e^+e^-$ events associated with the
$t\bar{t}$-pair production are calculated in
Section~\ref{sec:tt_event} in the framework of  perturbative QCD.
In Section~\ref{sec:n_num} the numerical estimations and our main
results are presented.

\section{Hadron multiplicity in $\mathbf{e^+e^-}$ annihilation \\
associated with $\mathbf{c\bar{c}}$ or $\mathbf{b\bar{b}}$-pair
production}
\label{sec:hadron_mult}

Hadron multiplicity  in $q\bar{q}$ event, $N_{q\bar{q}}(W)$, can
be represented in the following general form~\cite{Petrov:95}:
\begin{equation}\label{mult}
N_{q\bar{q}}(W) = 2 n_{q} + C_F \!\int\limits_{Q_0^2}^{W^2} \!
\frac{dk^2}{k^2} \,\frac{\alpha_s(k^2)}{\pi} \, n_g(k^2) \, E_q
\Big( \frac{k^2}{W^2} \Big) \;,
\end{equation}
where $q$ means a type of quarks produced in the process of
$e^+e^-$ annihilation into hadrons at the collision energy $W$. In
what follows, the notation $q=Q$ (\emph{heavy quark}) will mean
charm or beauty quark, while the notation $q=l$ (\emph{light
quark}) will correspond to a massless case (when a pair of $u,\ d$
or $s$-quarks is produced, whose masses are assumed to be equal to
zero). The top quark production ($q=t$) will be studied in
Sections~\ref{sec:tt_event} and \ref{sec:n_num}.

The first term in the r.h.s. of Eq.~\eqref{mult}, $2n_q$, is the
multiplicity of primary (anti)quark of the type $q$ (i.e. the
multiplicity from the leading hadron which contains this
(anti)quark). It is taken from an analysis of the data ($2n_c =
5.2$, $2n_b = 11.1$~\cite{Schumm:92}, and $2n_l =
2.4$~\cite{Chrin:94}).

The quantity $n_g(k^2)$ in \eqref{mult} is the mean multiplicity
of the gluon jet with a virtuality $k^2$, for which we will take a
QCD-based parametric form, with parameters fit to data, while
$E_q(k^2/W^2)$ is the inclusive spectrum of the gluon jet emitted
by primary quarks. It was explained in detail in
Ref.~\cite{Petrov:95} that one should not consider this mechanism
of hadron production via gluon jets as due to ``a single cascading
gluon''. That quantity $E(k^2/W^2)$ is an inclusive spectrum of
the gluon jets is seen, e.g., from the fact that the average
number of jets $\int dk^2\!/k^2 \,E_q(k^2/W^2) \neq 1$.

 $Q_0$ is a phenomenological parameter denoting the
scale at which ``preconfinement'' of the off-shell partons occurs
(as explained in Ref.~\cite{BCM}).

Let us introduce variables%
\footnote{We will often use ``rapidity-like'' variables (analogous
to $\eta$ and $Y$) instead of the energy variable $W$ throughout
the paper.}
\begin{equation}\label{eta}
\eta = \ln \frac{W^2}{k^2}
\end{equation}
and
\begin{equation}\label{Y}
Y = \ln \frac{W^2}{Q_0^2} \;,
\end{equation}
as well as notation
\begin{equation}\label{n_g_reduced}
\hat{n}_g = \frac{C_F \, \alpha_s(k^2)}{\pi} \, n_g(k^2) \;,
\end{equation}
where $C_F = (N_c^2 - 1)/2 N_c$, and $N_c=3$ is a number of
colors. Then Eq.~\eqref{mult} can be represented as
\begin{equation}\label{mult_Y}
N_{q\bar{q}}(Y) = 2\,n_{q} + \int\limits_0^Y \! d \eta \,
\hat{n}_g(Y - \eta) \, E_q(\eta) \equiv 2\,n_{q} +  N_q(Y) \;.
\end{equation}
In particular, $N_{l\bar{l}}(Y)$ means the multiplicity of hadrons
in light quark events, while $N_{Q\bar{Q}}(Y)$ denotes the
multiplicity of hadrons in a process when a pair of the heavy
quarks is produced.

The physical meaning of the function
\begin{equation}\label{cent_mult_Y}
N_q(Y) = \int\limits_0^Y \! d \eta \, \hat{n}_g(Y - \eta)E_q(\eta)
\equiv \int\limits_0^Y \! d \eta' \, \hat{n}_g(\eta') \, E_q(Y -
\eta') \;
\end{equation}
is the following. It describes the average number of hadrons
produced in virtual gluon jets emitted by the primary quark and
antiquark of the type $q$. In other words, it is the multiplicity
in $q\bar{q}$ event except for multiplicity of the decay products
of the primary quarks at the final stage of hadronization (the
terms $2n_q$ in \eqref{mult_Y}).

\begin{figure}[ht]
\epsfysize=6cm \epsffile{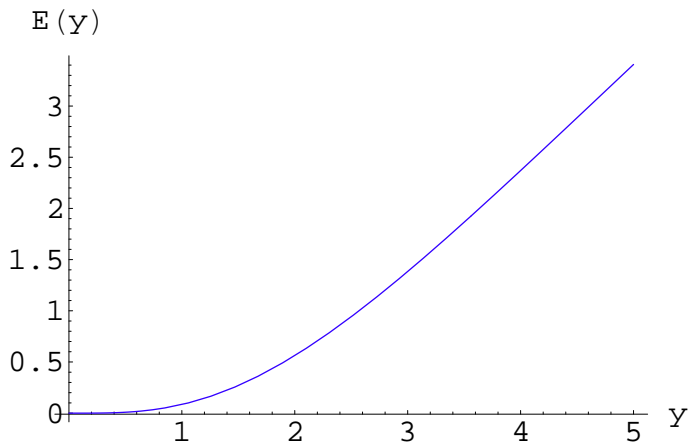}
\caption{The function $E(y)$.}
\label{fig:E}
\end{figure}

For the massless case, the function $E \equiv E_l$ was calculated
in our paper~\cite{Petrov:95}. In terms of variable
\begin{equation}\label{sigma_vs_eta}
\sigma = \exp (- \eta) \;,
\end{equation}
it looks as
\begin{eqnarray}\label{E}
E[\eta(\sigma)] &=& (1 + 2\sigma + 2\sigma^2) \ln \frac{1}{\sigma}
- \frac{3 + 7\sigma}{2} (1 - \sigma) - \sigma (1 + \sigma) \left(
\ln \frac{1}{\sigma} \right)^2
\nonumber \\
&+& 4\sigma(1 + \sigma) \Big[ \frac{\pi^2}{12} + \ln \sigma \ln (1
+ \sigma) + \mathrm{Li}_2(-\sigma) \Big] \;,
\end{eqnarray}
where $\mathrm{\rm Li}_2(z)$ is the Euler
dilogarithm. The function $E(\eta)$ is presented in
Fig.~\ref{fig:E}. It has the asymptotics
\begin{equation}\label{E_asym}
E(\eta)\Big|_{\eta \rightarrow \, \infty} = \eta - \frac{3}{2} \;.
\end{equation}

The derivative of $E(\eta)$ is positive, and $\partial
E(\eta)/\partial \eta = 0$ at $\eta=0$. As a result, the
associated multiplicity $N_q(W)$ \eqref{cent_mult_Y} is a
monotonic increasing function of the energy $W$ for any positive
function $n_g(k^2)$ since
\begin{equation}\label{derivative of_N}
\frac{\partial N_l(Y)}{\partial Y} = \int_0^Y \! d \eta \,
\hat{n}_g(\eta) \, \frac{\partial E(Y - \eta)}{\partial Y} \;.
\end{equation}

Now consider the multiplicity difference in events with the light
and heavy flavors ($Q=c$ or $b$):
\begin{equation}\label{delta_Ol_definition}
\delta_{Ql} = N_{Q\bar{Q}} - N_{l\bar{l}} \;.
\end{equation}
At $W \gg m_Q$, one can neglect small power-like corrections
$\mathrm{O}(m^2/W^2)$. In such a case, the quantity $\delta_{Ql}$
is defined by~\cite{Petrov:95}
\begin{equation}\label{delta_Ql}
\delta_{Ql} = 2(n_Q - n_l) - \Delta N_Q(Y_Q) \;,
\end{equation}
where the notation
\begin{equation}\label{delta_N}
\Delta N_Q(Y_Q) = N_l - N_Q = \int\limits_{-\infty}^{Y_Q} \! dy \,
\hat{n}_g (Y_Q - y) \, \Delta E_Q(y) \;,
\end{equation}
as well as variables
\begin{equation}\label{y}
y = \ln \frac{m_Q^2}{k^2}
\end{equation}
and
\begin{equation}\label{Y_Q}
Y_Q = \ln \frac{m_Q^2}{Q_0^2} \;
\end{equation}
are introduced. The lower limit of integration in
Eq.~\eqref{delta_N}, $-\ln (W^2/m_Q^2)$, is taken $-\infty$
because of the fast convergence of the integral at negative $y$.

\begin{figure}[ht]
\epsfysize=6cm \epsffile{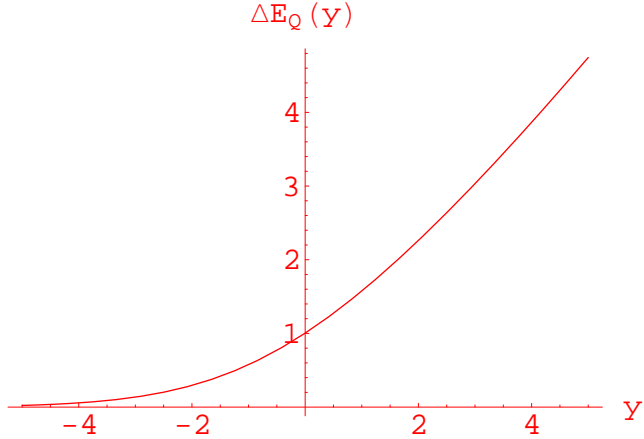}
\caption{The function $\Delta E_Q(y)$.}
\label{fig:deltaE_Q}
\end{figure}

Let us use another dimensionless variable
\begin{equation}\label{ro_vs_y}
\rho = \exp (-y) \;.
\end{equation}
The explicit form of $\Delta E_Q$ was derived in our
paper~\cite{Petrov:95} (see Fig.~\ref{fig:deltaE_Q}):
\begin{align}\label{Delta_E}
\Delta E_Q[y(\rho)] & = \Big[ 1 + \rho \, \Big( \frac{7}{2}\rho -
3 \Big) \ln \frac{1}{\rho} + \Big( \frac{9}{2} + 7\rho \Big)
\nonumber \\
& +  \rho \, (7\rho - 20) \, J(\rho) + 20 \, \frac{1 -
J(\rho)}{\rho - 4} \Big] \;.
\end{align}
where
\begin{equation}\label{J}
J(\rho) =
  \begin{cases}
     \sqrt{\frac{\rho}{\rho - 4}}
     \ln \left( \frac{\sqrt{\rho} \, + \sqrt{\rho - 4}}{2}
     \right),
     & \rho > 4 \;, \cr
     \ 1 \;, & \rho = 4 \;, \cr
     \sqrt{\frac{\rho}{4 - \rho}}
     \arctan \left( \frac{\sqrt{4 - \rho}}{\rho} \right),
     & \rho < 4 \; .
  \end{cases}
\end{equation}
The function $\Delta E_Q(y)$ decreases at $y \rightarrow -\infty$
($\rho \rightarrow \infty$) as
\begin{equation}\label{large_rho}
\Delta E_Q(y)\Big|_{y \rightarrow -\infty} \simeq \frac{11}{3} \,
e^{-|y|} \;,
\end{equation}
and has the following asymptotics at $y \rightarrow \infty$ ($\rho
\rightarrow 0$):
\begin{equation}\label{small_rho}
\Delta E_Q(y)\Big|_{y \rightarrow \infty} \simeq  y - \frac{1}{2}
\;.
\end{equation}

Thus, we get the relation between average multiplicities of
hadrons in $Q\bar{Q}$ and $l\bar{l}$ events:
\begin{equation}\label{QQ_ll_relation}
N_{Q\bar{Q}}(W) = N_{l\bar{l}}(W) - \Delta N_{Q}(m_Q) + 2 (n_Q -
n_l) \;,
\end{equation}
with the multiplicity difference $\Delta N_{Q}$ defined by
Eq.~\eqref{delta_N}.

Our calculations~\cite{Petrov:95} of the multiplicity differences
$\delta_{Ql} = N_{l\bar{l}} - N_{Q\bar{Q}}$ ($Q = b, c$) with the
use of formula \eqref{QQ_ll_relation} appeared to be in a good
agreement with the data. Recently we have reconsidered the QCD
upper limit on quantity $\delta_{bl}$~\cite{Kisselev:06} which
appeared to be very close to all present experimental data on
$\delta_{bl}$.%
\footnote{The data on $N_{ll}$ as well as on $\delta_{bl}$ at
different energies corrected for detector effects as well as for
initial state radiation were recently cited in
\cite{Dokshitzer:06}.}

\section{Hadron  multiplicity in $\mathbf{e^+e^-}$ annihilation \\
associated with $\mathbf{t\bar{t}}$-pair production}
\label{sec:tt_event}

The goal of this paper is to calculate $N_{t\bar{t}}^h$, the
average multiplicity of hadrons produced in $e^+e^-$ events with
the primary $t \bar{t}$ pair. We consider the case when the top
(antitop) decay mode is pure hadronic. As a byproduct, we will
calculate $n_t$, the hadron multiplicity of the on-shell top decay
products.

We will assume that the square of the matrix element of the
process $e^+e^- \rightarrow t^*\bar{t}^* \rightarrow X$ is
factorized as follows:
\begin{align}\label{matrix_element}
|M(e^+e^- \rightarrow t^*\bar{t}^* \rightarrow \mathrm{\
hadrons})|^2 &= |M(e^+e^- \rightarrow t^*\bar{t}^* \rightarrow
t\bar{t} + \mathrm{\ hadrons})|^2
\nonumber  \\
&\times |M(t \rightarrow \mathrm{\ hadrons})|^2
\nonumber  \\
&\times |M(\bar{t} \rightarrow \mathrm{\ hadrons})|^2 \;,
\end{align}
where $t^*(\bar{t}^*)$ denotes the virtual top quark (antiquark).

The factorization of the matrix element \eqref{matrix_element}
means that there is no significant space-time overlap in the decay
products of the on-shell $t$ and $\bar{t}$-quarks.%
\footnote{Note that the off-shell $t$ and $\bar{t}$-quarks
fragment into hadrons through the emission of the gluon jets in a
coherent way (the first term in the r.h.s of
Eq.~\eqref{matrix_element})}
The QCD non-singlet evolution of the primary virtual $t$-quark is
very slow because the difference of virtualities in logarithmic
scale is very small down to the top quark mass. In other words,
the virtual $t$-quark becomes ``real'' after just a few gluon
radiation.

The effect of possible color reconnection was investigated by
comparing hadronic multiplicities in $e^+e^- \rightarrow W^+W^-
\rightarrow q\bar{q}'q\bar{q}'$ and $e^+e^- \rightarrow W^+W^-
\rightarrow q\bar{q}'l\bar{\nu}_l$ events. No evidence for final
state interactions was found by measuring the difference $\langle
n^h_{4q} \rangle - 2\langle n^h_{2q l\bar{\nu}}
\rangle$~\cite{OPAL,DELPHI}. From the space-time point of view $W$
bosons and $t$-quarks behave in a similar way, i.e. the latter
manage to cover the distance $\Delta l \sim 1/\Gamma_t$, where
$\Gamma_t$ is the full width of the top. Since $\Gamma_t \simeq
\Gamma_W$, we expect no interference effects in the decays of the
on-shell $t$ and $\bar{t}$-quarks.

According to Eq.~\eqref{matrix_element}, the associative
multiplicity in $t\bar{t}$-event is given by the formula:
\begin{equation}\label{tt_ll_relation}
N_{t\bar{t}}^h(W, m_t) = N_{t}(W, m_t) + 2 n_t  \;,
\end{equation}
where
\begin{equation}\label{N_t}
N_{t}(W, m_t) =  C_F \int\limits_{Q_0^2}^{(W-2m_t)^2} \!
\frac{dk^2}{k^2} \,\frac{\alpha_s(k^2)}{\pi} \, n_g(k^2) \, E_t
(W^2, k^2, m_t^2) \;.
\end{equation}

Here and in what follows we will assume that the collision energy
is a typical ILC energy, $W = 500$ GeV, for definiteness. In such
a case, contrary to Eq.~\eqref{cent_mult_Y}, power corrections
$\mathrm{O}(m_t/W)$ should be taken into account. The explicit
form of the inclusive distribution of the gluon jets with the
invariant mass $\sqrt{k^2}$ looks like
\begin{align}\label{E_t}
E_t (q^2, k^2, m^2) &= \left( -k \frac{\partial}{\partial k}
\right) \int\limits_1^A d\eta \left\{ \left[ \frac{1}{\eta}
\frac{(q^2 + k^2)^2 - 4m^4}{q^4} - 2 \, \frac{k}{q} \, \frac{q^2 +
k^2 + 2m^2}{q^2} \right. \right.
\nonumber \\
&+ \left. 2\eta \, \frac{k^2}{q^2} \, \right] \ln \left[
\frac{\eta + \sqrt{\eta^2 - 1} \sqrt{(A - \eta)/(A_0 -
\eta)}}{\eta - \sqrt{\eta^2 - 1} \sqrt{(A - \eta)/(A_0 - \eta)}}
\right]
\nonumber \\
&- 2\,\frac{k^2}{q^2} \, \sqrt{\eta^2 - 1} \frac{ \sqrt{(A -
\eta)}}{ \sqrt{(A_0 - \eta)}}
\nonumber \\
& \times \left. \left[ 1 + \frac{(1 +
2m^2/q^2)(1+2m^2/k^2)}{\eta^2 - (\eta^2 - 1) (A - \eta)/(A_0 -
\eta) } \right] \right\} \;,
\end{align}
where $k \equiv \sqrt{k^2}$, $q \equiv \sqrt{q^2}$, and the
following notations are introduced:
\begin{equation}\label{A_A0}
A = \frac{q^2 + k^2 - 4m^2}{2 q k}, \qquad A_0 = \frac{q^2 +
k^2}{2 q k} \;.
\end{equation}

This formula has been derived by calculating QCD diagrams in the
first order in the strong coupling constant.%
\footnote{See our comments after Eq.~\eqref{mult}.}
In the massless case ($m=0$), we immediately come to the function
$E(k^2/q^2)$~\eqref{E}, while by neglecting small corrections
$\mathrm{O}(m^2/q^2)$, one can derive%
\footnote{After variables are properly changed.}
the explicit form of the function $\Delta E_Q(k^2/m^2) = E_Q(q^2,
k^2, m^2) - E(q^2, k^2)$~\eqref{Delta_E}. In our case ($q^2 =
W^2$, $m=m_t$) we will estimate the integrals in Eqs.~\eqref{N_t},
\eqref{E_t} numerically (for details, see
Section~\ref{sec:n_num}).

\begin{figure}[ht]
\begin{center}
\epsfysize=3cm \epsffile{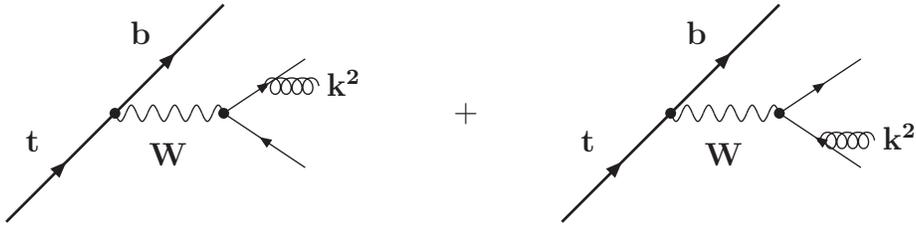} \caption{The emission
of the gluon jets (spiral lines) by the quark pair resulting from
the decay of the $W^+$-boson in the first order in the strong
coupling constant. The $W^+$ boson is produced in the weak decay
of the top.} \label{fig:gluon_W_boson}
\end{center}
\end{figure}

Now let us calculate another quantity in
Eq.~\eqref{tt_ll_relation}, $n_t$, which describes the hadronic
multiplicity of the $t$-quark decay products. The top weakly
decays into $W^{+}$ boson and $b$-quark. In its turn, the $W^{+}$
boson decays into a quark-antiquark pair.%
\footnote{Remember that we are interested in hadronic decays of
the $W$ boson.}
The quark-antiquark system results in massive jets which fragment
into hadrons (see Fig.~\ref{fig:gluon_W_boson}).

\begin{figure}[ht]
\begin{center}
\epsfysize=3cm \epsffile{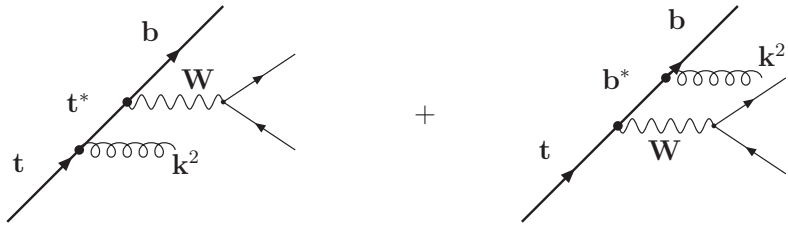} \caption{The emission of
the gluon jets by the on-shell top quark and by off-shell bottom
quark in the first order in the strong coupling constant. The
emissions take place before and after the weak decay of the top,
respectively. Off-shell quarks are denoted as $t^*$ and $b^*$.}
\label{fig:gluon_tb}
\end{center}
\end{figure}

The gluon jets can be also emitted by the on-shell $t$-quark
before its weak decay (the first diagram in
Fig.~\ref{fig:gluon_tb}) or by off-shell bottom quark (the second
diagram in Fig.~\ref{fig:gluon_tb}).
At the end of these emissions, the on-shell $b$-quark weakly
decays into hadrons whose average multiplicity is equal to $n_b$.
Since $W$-boson is a colorless particle, the diagrams in
Fig.~\ref{fig:gluon_tb} do not interfere with those presented in
Fig.~\ref{fig:gluon_W_boson}.

Thus, the multiplicity $n_t$ is a sum of three terms:
\begin{equation}\label{n_t}
n_t = n_W + n_{tb} + n_b \;.
\end{equation}
The quantity $n_b$ is experimentally measurable
one~\cite{Schumm:92}. The first term in Eq.~\eqref{n_t}, $n_W$, is
the hadron multiplicity of the $W$ boson decay products. The
second term, $n_{tb}$, is the hadron multiplicity in the gluon
jets emitted by the on-shell top quark before its weak decay as
well as by the bottom quark after the top decay.

\subsection{Multiplicity of W boson decay products}
\label{subsec:n_w}

The $W^+$ boson can decay either into two light quarks ($u\bar{d}$
and $u\bar{s}$ pairs) or into $c\bar{d}$ ($c\bar{s}$) pair. The
former case is treated analogously to the light quark event in
$e^+e^-$ annihilation taken at the collision energy $W=m_W$. Here
we will study the latter case.

Let $N_{Ql}(W)$ be hadronic multiplicity associated with the
production of \emph{one} heavy quark (antiquark) of the type $Q$
and \emph{one} light antiquark (quark) of the type $l$:
\begin{equation}\label{mult_Ql}
N_{Ql}(W)  =  (n_Q + n_l) + \hat{N}_{Ql}(W) \;.
\end{equation}
Now let us introduce the notation (not to confuse with $\Delta
N_{Q}$ from above):
\begin{equation}{}
\Delta N_{Ql} = N_l - \hat{N}_{Ql} \;.
\end{equation}
Then the first term in the r.h.s. of Eq.~\eqref{n_t} is given in
terms of the function $\Delta N_{cl}$ by the formula:
\begin{equation}\label{n_t_W}
n_W= N_{l\bar{l}}(Y_W) + \frac{1}{2} \, \left[ - \Delta
N_{cl}(Y_c) + n_c - n_l \right] \;,
\end{equation}
where $Y_c = \ln (m_c^2/Q_0^2)$ and
\begin{equation}\label{Y_W}
Y_W = \ln \frac{m_W^2}{Q_0^2} \;.
\end{equation}
The function $ N_{l\bar{l}}(Y)$ in \eqref{n_t_W} is the hadronic
multiplicity in light quark events.

Thus, we need to find an expression for $\Delta N_{cl}$. Note that
the formulae \eqref{delta_N}, \eqref{Delta_E} from
Section~\ref{sec:hadron_mult} correspond to the case when \emph{a
pair} of heavy or \emph{pair} of light quarks is produced. Now we
have to study the case when hadrons are produced in association
with \emph{a single} heavy quark (namely, $c$-quark) and one light
quark.

\begin{figure}[ht]
\epsfysize=6cm \epsffile{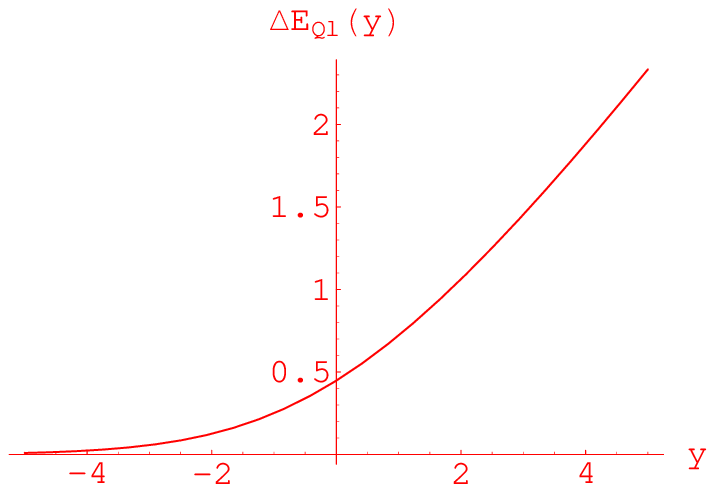}
\caption{The function $\Delta E_{Ql}(y)$.}
\label{fig:deltaE_Ql}
\end{figure}

Our QCD calculations result in the following representation for
the multiplicity difference (see Appendix for details):
\begin{equation}\label{delta_N_Ql}
\Delta N_{Ql}(Y_Q) = \int\limits_{-\infty}^{Y_Q} \! dy \,
\hat{n}_g (Y_Q - y) \, \Delta E_{Ql}(y) \;,
\end{equation}
with the dimensionless function $\Delta E_{Ql}$:
\begin{align}\label{delta_E_Ql}
\Delta E_{Ql}[y(\rho)] & = \frac{1}{4} \, [2 + \rho \, (3\rho -
2)] \ln \frac{1}{\rho} +  \frac{1}{4} \, (5 + 6\rho)
\nonumber \\
& +  \frac{1}{2} \, \rho \, (3\rho - 8) \, J(\rho) + 6 \, \frac{1
- J(\rho)}{\rho - 4} \;.
\end{align}
Here $\rho = \exp(-y)$. The quantity $J(\rho)$ was defined above
\eqref{J}. The function $\Delta E_{Ql}(y)$ is shown in
Fig.~\ref{fig:deltaE_Ql}.

Since
\begin{equation}\label{large_negative_y}
\Delta E_{Ql}(y)\Big|_{y \rightarrow  - \infty} \simeq \frac{3}{2}
\, e^{-|y|}\;,
\end{equation}
the integral \eqref{delta_N_Ql} converges rapidly at the lower
limit. Asymptotics of $\Delta E_{Ql}(y)$ at large $y$ is the
following:
\begin{equation}\label{large_positive_y}
\Delta E_{Ql} (y)\Big|_{y \rightarrow \infty} \simeq \frac{1}{2}
\, \Big( y - \frac{1}{2}  \Big) \;.
\end{equation}

\begin{figure}[ht]
\epsfysize=6cm \epsffile{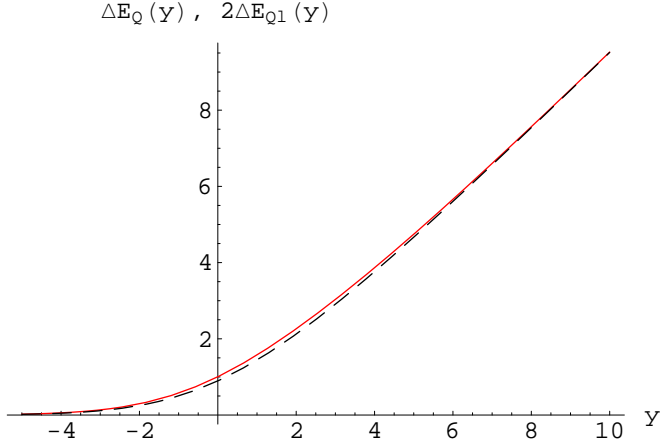}
\caption{The function $\Delta E_Q(y)$ (solid line) vs. function
$2\Delta E_{Ql}(y)$ (dashed line).}
\label{fig:dE_Q_dE_Ql}
\end{figure}

We derive from Eqs.~\eqref{small_rho}, \eqref{large_positive_y}
that $\Delta E_{Ql}(y) = 0.5 \, \Delta E_Q(y)$ at large $y$.
Numerical calculations show that $2\Delta E_{Ql}$ is very close to
$\Delta E_Q$ at all $y$ (see Fig.~\ref{fig:dE_Q_dE_Ql}). Thus, we
can put for our further numerical estimates:
\begin{equation}\label{N_cl_vs_N_c}
\Delta N_{cl} = \frac{1}{2} \, \Delta N_{c} \;.
\end{equation}
This relation means that
\begin{equation}\label{Ql_vs_ll_QQ}
N_{Q\bar{l}} = \frac{1}{2} \, [N_{l\bar{l}} + N_{Q\bar{Q}}]  =
N_{l\bar{l}} + \frac{1}{2} \, \delta_{Ql}\;.
\end{equation}
Correspondingly, we obtain:
\begin{equation}\label{n_W}
n_W= N_{l\bar{l}}(m_W) + \frac{1}{4} \, \delta_{cl} \;.
\end{equation}

\subsection{Multiplicity of top and bottom decay products}
\label{subsec:n_tb}

As was already said above, the on-shell top quark can emit jets
before it weakly decays into $W^{+}b$. After the weak decay of the
top, the off-shell $b$-quark ``throws off'' its virtuality by
emitting massive gluon jets. The fragmentation of these massive
gluon jets into hadrons results in the average hadron multiplicity
$n_{tb}$.

To calculate the multiplicity $n_{tb}$, one has to derive the
inclusive spectrum of the gluon jets, emitted by the top and
bottom quarks. Let us denote it as $E_{tb}$. Then the multiplicity
$n_{tb}$ will be given by the formula:
\begin{equation}\label{n_tb}
n_{tb} =  \int\limits_{0}^{Y_{tb}} \! dy \, \hat{n}_g (Y_{tb} - y)
\, E_{tb}(y) \;,
\end{equation}
where
\begin{equation}\label{y_tb}
y = \ln \frac{(m_t-m_W-m_b)^2}{k^2} \;,
\end{equation}
and
\begin{equation}\label{hat_Y_tb}
Y_{tb} = \ln \frac{(m_t-m_W-m_b)^2}{Q_0^2} \;,
\end{equation}
with $k^2$ being the  gluon jet invariant mass, $(m_t-m_W-m_b)^2$
its upper bound.

\begin{figure}[ht]
\epsfysize=6cm \epsffile{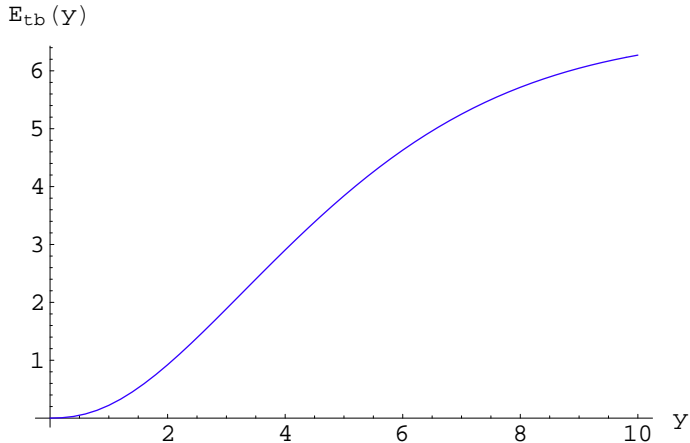}
\caption{The function $E_{tb}(y)$.}
\label{fig:E_tb}
\end{figure}

In the lowest order in the strong coupling constant, the quantity
$E_{tb}(y)$ is given by two diagrams in Fig.~\ref{fig:gluon_tb}.
It is presented by an integral which depends on the ratio
$k^2/m_t^2$, as well as on mass ratios $m_W^2/m_t^2$ and
$m_b^2/m_t^2$. This integral cannot be calculated analytically,
but can be estimated numerically. The function $E_{tb}(y)$ is
presented in Fig.~\ref{fig:E_tb}. It is worth to note that in the
Feynman gauge the dominating contribution to $E_{tb}(y)$ comes
from the \emph{interference} of two diagrams shown in
Fig.~\ref{fig:gluon_tb}.

\subsection{Associated multiplicity of hadrons in
$\mathbf{t\bar{t}}$ events}
\label{subsec:n_tt}

The formulae of the previous subsections enable us to derive the
average multiplicity of the charged hadrons in $e^+e^-$
annihilation at the collision energy $W$ associated with the
production of the $t\bar{t}$-pair. It is of the form:
\begin{align}\label{mult_tt_final}
N_{t\bar{t}}^h(W,m_t) & = N_t(W, m_t) + 2[ N_{l\bar{l}}(m_W)+
n_{tb} + n_b ]
\nonumber \\
& + [ - \Delta N_{cl}(m_c) + n_c - n_l ] \;.
\end{align}

Let us remind to the reader the meaning of all quantities in
Eq.~\eqref{mult_tt_final}. The function $N_t(W, m_t)$ describes
the average number of hadrons produced in association with the
$t\bar{t}$-primary pair, \emph{except for} the decay products of
the top(antitop)~\eqref{N_t}. The quantity $N_{l\bar{l}}(m_W)$ is
the mean hadron multiplicity in the light quark event taken at the
energy $E = m_W$. The hadron multiplicity $n_{tb}$ comes from the
emission by $t$ and $b$ quarks \eqref{n_tb}. The quantity $n_l$ is
the mean multiplicity of hadrons produced in the decay of the
on-shell primary quark $q$ ($q = l, \, c,\,b$). Finally, the
combination $[\Delta N_{cl}(m_c) + n_l - n_c$] is the difference
of multiplicities in the processes with the primary $l\bar{l}$-
and $c\bar{l}$-pairs. As for the hadron multiplicity resulting
from the decay of the on-shell top (anti)quark, it is given by
\begin{align}\label{n_t_final}
n_t &= n_W + n_{tb} + n_b
\nonumber \\
&= N_{l\bar{l}}(m_W)  + n_{tb} + n_b + \frac{1}{2} \, [- \Delta
N_{cl}(m_c) + n_c - n_l] \;.
\end{align}
The expressions for $N_{l\bar{l}}$, $\Delta N_{cl}$ are given by
Eqs.~\eqref{mult_Y}, \eqref{delta_N_Ql}, respectively.

Let us stress that $N_{l\bar{l}}$ and $n_q$ ($q=b, c, l$) are
extracted from the data, and $\Delta N_{cl}$ can be related with
the measurable quantities (see formulae
\eqref{N_cl_vs_N_c}-\eqref{n_W} in the end of
subsection~\ref{subsec:n_w}):
\begin{equation}\label{delta_N_cl_vs_delta_cl}
\Delta N_{cl} \simeq \frac{1}{2} \, \Delta N_{c} = n_c - n_l -
\frac{1}{2} \, \delta_{cl} \;.
\end{equation}
Then we obtain:
\begin{equation}
n_t = N_{l\bar{l}}(m_W) + \frac{1}{4} \, \delta_{cl} + n_{tb} +
n_b \;, \label{n_t_appr}
\end{equation}
and
\begin{equation}
N_{t\bar{t}}^h(W,m_t) = N_t(W, m_t) + 2 \Big[ N_{l\bar{l}}(m_W) +
\frac{1}{4} \, \delta_{cl} + n_{tb} + n_b \, \Big] \;,
\label{mult_tt_appr}
\end{equation}
where $n_{tb}$ is defined above \eqref{n_tb}. In what follows, we
will use the value
\begin{equation}\label{delta_cl_exp}
\delta_{cl}= 1.03 \pm 0.34
\end{equation}
from Ref.~\cite{Dokshitzer:06}.

\section{Numerical estimates of hadron multiplicities.}
\label{sec:n_num}

\begin{figure}[ht]
\begin{center}
\epsfysize=7cm \epsffile{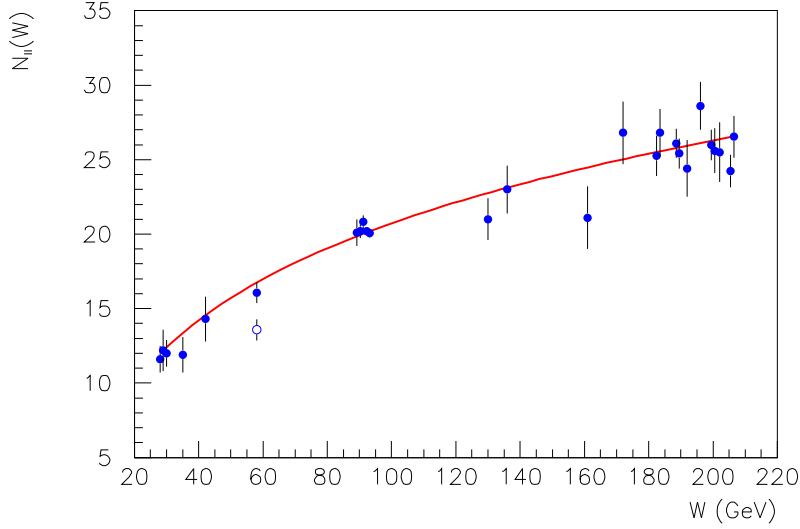}
\end{center}
\caption{The fit of the data on the hadron multiplicity in light
quark events (solid line, see formula in the text). The data are
taken from Ref.~\cite{Dokshitzer:06}.}
\label{fig:light_mult}
\end{figure}

In order to estimate the multiplicity of the decay products of the
top (formula~\eqref{n_t_appr}), one has to know the energy
dependence of the hadron multiplicity in the light quark event.
The latter is defined by Eq.~\eqref{mult_Y}, where the function
$\hat{n}_g$ is related with gluon jet multiplicity
$n_g(k^2)$~\eqref{n_g_reduced}. We have fitted the data on
$N_{l\bar{l}}(W)$ by using the following QCD-motivated expression
for  $n_g(k^2)$:
\begin{equation}\label{fit}
n_g(k^2) = a + b \exp {\Big[ c\sqrt{\ln (k^2/Q_0^2)} \Big] } \;,
\end{equation}
where $c=1.63$, and  $k^2$ is the invariant mass of the jet. We
have got the following values of the parameters:
\begin{equation}\label{fit_parameters}
a = 3.89, \qquad b = 0.01, \qquad Q_0 = 0.87 \mathrm{\ GeV} \;.
\end{equation}
The result of our fit is presented in Fig.~\ref{fig:light_mult} in
comparison with the data.
Note that $\chi^2/d.o.f.$ becomes twice smaller if one eliminates
the experimental point at $W=58$ GeV (open circle in
Fig.~\ref{fig:light_mult}), which lies much lower than neighboring
points, is crossed out from the fit. In such a case, the values of
the parameters are practically the same as in
\eqref{fit_parameters} with $\chi^2/d.o.f. = 0.90$.

For our numerical estimates we shall use $m_W = 80.40 \pm 0.03$
GeV~\cite{PDG} and the recent value of the top mass~\cite{top}:
\begin{equation}\label{top_mass}
m_t = 170.9 \pm 1.8 \mathrm{\ GeV} \;.
\end{equation}
As for the bottom quark, its pole mass is quoted in \cite{PDG} to
be $m_b = 4.7-5.0$ GeV. We will take the average value
\begin{equation}\label{bottom_mass}
m_b = 4.85 \pm 0.15 \mathrm{\ GeV} \;.
\end{equation}
By using our fit, we obtain $N_{l\bar{l}}(m_W) = 19.09$. Then we
get (see Eqs.~\eqref{n_W}, \eqref{delta_cl_exp}, \eqref{n_tb}):
\begin{align}
n_W &= 19.34 \pm 0.10 \;, \label{n_W_num} \\
n_{tb} &= 16.14 \pm 0.24 \;. \label{n_tb_num}
\end{align}
The error in Eq.~\eqref{n_W_num} is defined by that of the
multiplicity difference $\delta_{cl}$~\eqref{delta_cl_exp}, while
that in  Eq.~\eqref{n_tb_num} comes from uncertainties of the
quark masses $m_t$~\eqref{top_mass} and $m_b$~\eqref{bottom_mass}.

Our result \eqref{n_W_num} is in a nice agreement with the
experimental values from Ref.~\cite{OPAL},
\begin{align}
n_W &= 19.3 \pm 0.3 \pm 0.3 \;, \label{OPAL}
\intertext{and Ref.~\cite{DELPHI},}
n_W &= 19.44 \pm 0.13 \pm 0.12 \;. \label{DELPHI}
\end{align}

Now let us calculate the associated hadron multiplicity in
$t\bar{t}$ event \eqref{mult_tt_appr} at fixed energy $W = 500$
GeV. To do this, we need to estimate the multiplicity $N_t(W,
m_t)$ by using formulae \eqref{N_t} and \eqref{E_t}:
\begin{equation}\label{N_t_num}
N_t(W = 500 \mathrm{\ GeV}) = 4.61 \pm 0.11 \;.
\end{equation}
The errors in \eqref{N_t_num} come from top quark mass errors. It
follows From Eqs.~\eqref{n_t_num}, \eqref{N_t_num} that
$N_{t\bar{t}}(e^+e^- \rightarrow t\bar{t} \rightarrow \mathrm{\
hadrons}) = 86.67 \pm 0.55$.

In order to estimate possible theoretical uncertainties, we have
repeated our calculations taking the different form of the average
multiplicity of the gluon jet (compare with the QCD-based
expression \eqref{fit}):
\begin{equation}\label{fit_modified}
\hat{n}_g(k^2) = A + B \ln^2 \frac{k^2}{Q_0^2} \;.
\end{equation}
It appeared that the data on the average multiplicity in light
quark events can be fitted well by using this expression (with $A
= 4.21$, $B = 0.012$ and $Q_0= 0.93$ GeV). In particular, we have
obtained the following average values for the hadronic
multiplicities: $n_W = 19.52$, $n_{tb} = 16.43$, $N_t = 4.59$.
Thus, theoretical uncertainties can be estimated to be $0.47$ and
$0.96$ for $n_t$ and $N_{t\bar{t}}$, respectively.

Taking into account the phenomenological value of
$n_b$~\cite{Dokshitzer:06},
\begin{equation}\label{n_b_exp}
n_b = 5.55 \pm 0.09 \;,
\end{equation}
we obtain from \eqref{n_t_appr}:
\begin{equation}\label{n_t_num}
n_t(t \rightarrow \mathrm{\ hadrons})  = 41.03 \pm 0.54 \;.
\end{equation}
In the case when the $W$ boson decays into leptons, we predict:
\begin{equation}\label{n_t_semi_lepton}
n_t(t \rightarrow l \bar{\nu_l}  + \mathrm{\ hadrons}) = 21.69 \pm
0.53 \;.
\end{equation}

As a result, we obtain the average hadron multiplicity in $e^+e^-$
annihilation with the primary $t\bar{t}$-pair:
\begin{equation}\label{N_tt_X}
N_{t\bar{t}}(e^+e^- \rightarrow t\bar{t} \rightarrow \mathrm{\
hadrons}) = 86.67 \pm 1.11 \;.
\end{equation}
In the case when both $W$ bosons decay into leptons, the mean
multiplicity of the hadrons is expected to be
\begin{equation}\label{N_tt_WW_X}
N_{t\bar{t}}(e^+e^- \rightarrow t\bar{t} \rightarrow W^+W^- +
\mathrm{\ hadrons}) = 47.99 \pm 0.59 \;.
\end{equation}
Finally, we predict that
\begin{equation}\label{N_tt_WW_bb_X}
N_{t\bar{t}}(e^+e^- \rightarrow t\bar{t} \rightarrow b\bar{b} \
W^+W^- + \mathrm{\ hadrons}) = 36.89 \pm 0.56 \;.
\end{equation}
All estimations \eqref{N_tt_X}-\eqref{N_tt_WW_bb_X} correspond to
the collision energy $W = 500$ GeV.  We can mention the estimation
of the hadron multiplicity from Ref.~\cite{Ballestrero:96},
$N_{t\bar{t}}(e^+e^- \rightarrow t\bar{t} \rightarrow b\bar{b} \
W^+W^- + \mathrm{\ hadrons}) \approx 29$, which was obtained for
$W = 390$ GeV and $m_t = 175$ GeV. For the same values of $W$ and
$m_t$, our formulae give $N_{t\bar{t}}(e^+e^- \rightarrow t\bar{t}
\rightarrow b\bar{b} \ W^+W^- + \mathrm{\ hadrons}) = 34.3$.

The formulae \eqref{n_t_num}-\eqref{N_tt_WW_bb_X} is our main
result. We hope that the hadron multiplicities of the top decay
products (Eqs.~\eqref{n_t_num} and  \eqref{n_t_semi_lepton}) will
be measured at the LHC.


\section*{Acknowledgements}

We are thankful to the referee for his comments and critical
remarks that helped us to improve the presentation of some our
results.


\setcounter{equation}{0}
\renewcommand{\theequation}{A.\arabic{equation}}

\section*{Appendix A}

Here we present some formulae for the case when the $W$ boson
decays into hadrons via production of $c\bar{s}$ (or $c\bar{d}$)
pair. Since the total width of the $W$ boson, $\Gamma_W$, is much
less than its mass, and the hadron multiplicity is a smooth
function of energy, we will use zero width approximation and take
the multiplicity at $W=m_W$. It can be shown that the account of
the $W$ boson width results in corrections which are numerically
small (less than $1.6 \%$, see Appendix~B).

\begin{figure}[ht]
\begin{center}
\epsfysize=5cm \epsffile{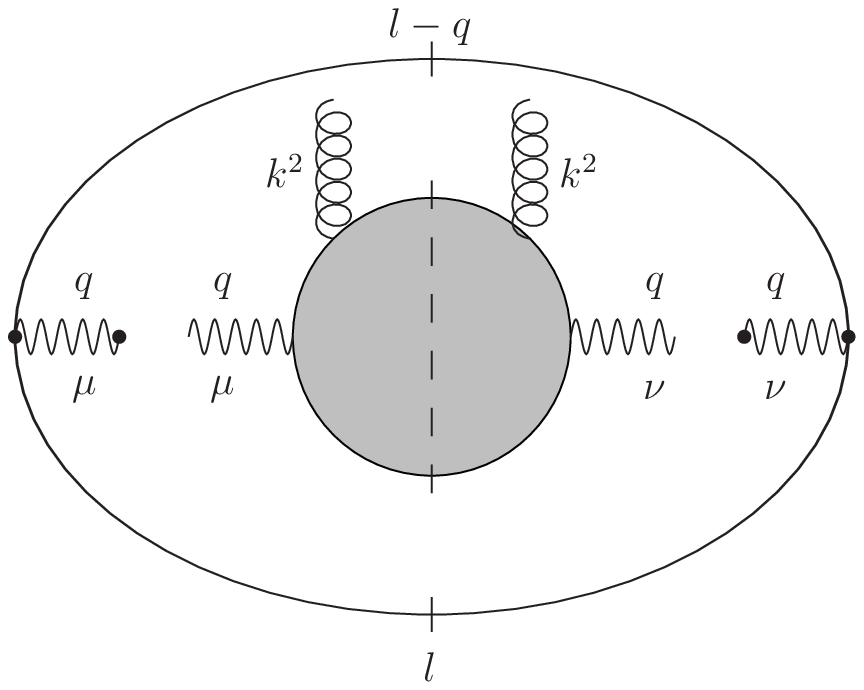}
\caption{The generalized diagram describing the inclusive spectrum
of the gluon jets (curly line) with the virtuality $k^2$ inside
the $W$ boson (wavy line). In its turn, the $W$ boson is a product
of the weak decay of the top quark (solid line with 4-momentum
$l$).}
\label{fig:A1}
\end{center}
\end{figure}

To calculate the inclusive spectrum of the gluon jets emitted by
the decay products of the $W$-boson, we need to calculate two
sub-diagrams of the diagram presented in Fig.~\ref{fig:A1}.

The exterior part of the diagram in Fig.~\ref{fig:A1} describes
the emission of the $W$ boson by the top quark, with $l$ being a
4-momentum of the $t$-quark, $q$ is a 4-momentum of the $W$ boson,
and $(l-q) $ is a 4-momentum of the $b$-quark. The corresponding
expression for this part of the diagram (after convolution with
the tensor parts of the $W$ boson propagators) looks like
\begin{align}\label{partonometer_W}
& \Pi_{\mu \nu}(l, \, q)  = 4 \Big\{ \, 4 \, l_{\mu} l_{\nu} - 2
\, (l_{\mu} q_{\nu} + q_{\mu} l_{\nu})  - g_{\mu \nu} (m_t^2 +
m_b^2 - m_W^2)
\nonumber \\
& - \frac{2}{m_W^2} \, [(m_t^2 - m_b^2)l_{\mu} - m_t^2 q_{\mu}] \,
q_{\nu}
 - \frac{2}{m_W^2} \, q_{\mu} \, [(m_t^2 - m_b^2)l_{\nu} - m_t^2
q_{\nu}]
\nonumber \\
& + \frac{1}{m_W^4} \, q_{\mu} q_{\nu} \, [(m_t^2 - m_b^2)^2 -
q^2(m_t^2 + m_b^2)] + 2 i \, \varepsilon_{\mu \nu \rho \sigma} \,
l^{\rho} q^{\sigma} \Big\} \;,
\end{align}
where $m_t$ and $m_b$ are masses of the top and beauty quark,
respectively. In what follows, we will neglect power corrections
of the type $\mathrm{O}(m_c/m_t)$ and  $\mathrm{O}(m_b/m_t)$.

The inner part of the diagram in Fig.~\ref{fig:A1} describes the
distribution of the massive gluon  jet with the invariant mass
$k^2$ produced by the $W$ boson. Let $D^{\mu \nu}$ be the
expression corresponding to this diagram. In the first order in
the strong coupling constant, $D^{\mu \nu}$ is represented by the
sum of three QCD diagrams presented in Figs.~\ref{fig:A2a},
\ref{fig:A2b} and \ref{fig:A2c}.
{
\newcounter{subfigure}
\renewcommand{\thefigure}{\arabic{figure}\alph{subfigure}}

\setcounter{subfigure}{1}

\begin{figure}[ht]
\begin{center}
\epsfysize=4cm \epsffile{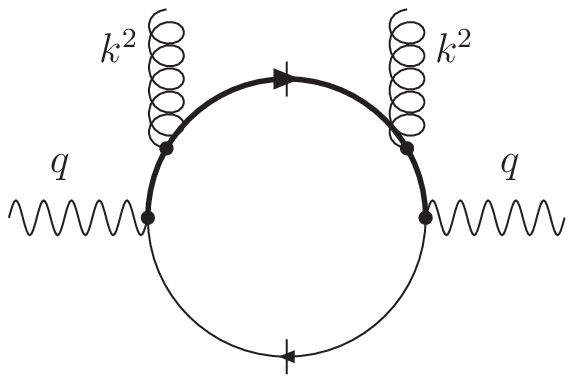}
\end{center}
\caption{The inclusive distribution of the massive gluon jet with
the virtuality $k^2$. The wavy line is the $W$ boson, whose
4-momentum is $q$. The thick quark line is a heavy quark, while
the thin line is a light quark. The cut quark lines mean that
these quarks are on-shell quarks.}
\label{fig:A2a}
\end{figure}

\addtocounter{figure}{-1}
\setcounter{subfigure}{2}

\begin{figure}[ht]
\begin{center}
\epsfysize=4cm \epsffile{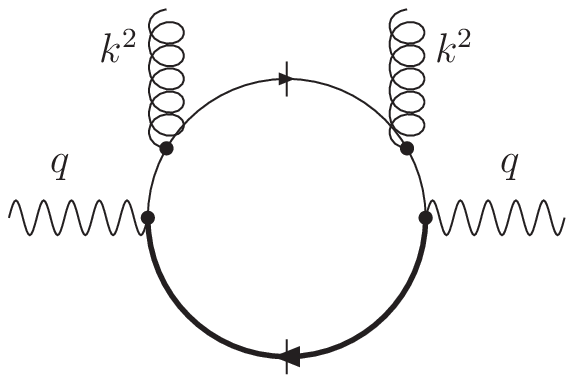}
\end{center}
\caption{The same as in Figs.~\ref{fig:A2a}, but with the gluon
jet emitted by the light quark.}
\label{fig:A2b}
\end{figure}

\addtocounter{figure}{-1}
\setcounter{subfigure}{3}

\begin{figure}[ht]
\begin{center}
\epsfysize=4cm \epsffile{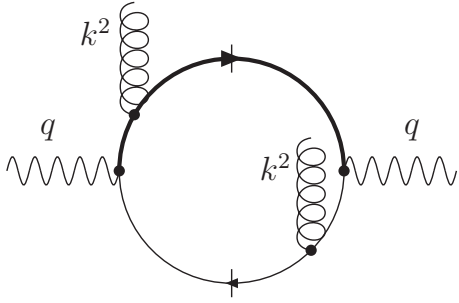}
\end{center}
\caption{The interference diagram which also contributes to the
inclusive distribution of the gluon jets with the virtuality $k^2$
inside the $W$ boson. The diagram are taken in the sum of the
diagrams with the factor 2.}
\label{fig:A2c}
\end{figure}
}

Since $q^{\mu} \Pi_{\mu \nu} = 0$, one needs to calculate only two
tensor structures in $D_{\mu \nu}$, namely, $g_{\mu \nu}$ and
$k_{\mu} k_{\nu}$. The convolution of the tensor $\Pi_{\mu \nu}$
with the tensor $D^{\mu \nu} = g^{\mu \nu} D_1 + k^{\mu} k^{\nu}
D_2 + \cdots$,
\begin{equation}\label{A}
A = \Pi_{\mu \nu} D^{\mu \nu} = g^{\mu \nu} \, \Pi_{\mu \nu}
D_1(k^2, qk) + k^{\mu} k^{\nu} \, \Pi_{\mu \nu} D_2(k^2, qk) \;,
\end{equation}
depends on Lorentz-invariant variables $k^2, qk$, $lk$. Moreover,
it is a polynomial of the second order in variable $lk$. One can
use the following useful relation:
\begin{equation}\label{change_variables_W}
\int \! \frac{d^4k}{(2\pi)^3} \, A(k^2, qk, lk) = \frac{1}{(2
\pi)^2 m_t^2 (1 - r)} \iiint \! d k^2 \, d (qk) \, d (lk) \,
A(k^2, qk, lk) \;,
\end{equation}
where
\begin{equation}\label{r}
r = \frac{m_W^2}{m_t^2} \;.
\end{equation}

It is naturally to integrate the function $A(k^2, qk, lk)$ first
in variable $lk$, whose lower and upper limits are
\begin{align}\label{int_limits_lk_W}
(lk)_{\pm} = \frac{1}{2r} \, \Big[ (qk)(1 + r) \pm (1 - r) \,
\sqrt{(qk)^2 - m_W^2 k^2} \Big] \;,
\end{align}
by using the following formulae:
\begin{align}\label{int_lk_W}
\int_{(lk)_{-}}^{(lk)_{+}} d (lk) & = \frac{1}{r} \, (1 - r) \,
\sqrt{(qk)^2 - m_W^2 k^2} \;,
\nonumber \\
\int_{(lk)_{-}}^{(lk)_{+}}  (lk) \, d (lk) & = \frac{1}{2r^2} \,
(1 - r^2) \, (qk) \, \sqrt{(qk)^2 - m_W^2 k^2} \;,
\nonumber \\
\int_{(lk)_{-}}^{(lk)_{+}}  (lk)^2 \, d (lk) & = \frac{1}{12r^3}
\, (1 - r) \, \Big[ 4(qk)^2 (1 + r + r^2) - m_W^2 k^2 (1 - r)^2
\Big]
\nonumber \\
& \times \sqrt{(qk)^2 - m_W^2 k^2} \;.
\end{align}

Note that the parts of tensors $\Pi_{\mu \nu}$ and $D^{\mu \nu}$
antisymmetric in indices%
\footnote{As the last term in Eq.~\eqref{partonometer_W}.}
give no contribution after integration in $lk$.

Integration limits in variable $qk$ looks like
\begin{align}\label{int_limits_qk_W}
(qk)_{-} & = \sqrt{m_W^2 k^2} \;,
\nonumber \\
(qk)_{+} & = \frac{1}{2} \, (m_W^2 + k^2) \;.
\end{align}

As a result, we obtain the hadron multiplicity associated with the
charm quark in a hadronic decay of the $W$ boson (with the
multiplicity of primary quark decay products subtracted):
\begin{align}\label{mult_cl_repres}
\hat{N}_{cl}  = \frac{1}{(2 \pi)^2 m_t^2 (1 - r) H} &
\int\limits_{Q_0^2}^{m_W^2} \! dk^2 \Big[ \frac{\partial}{\partial
k^2} n_g(k^2) \Big] \int\limits_{(qk)_{-}}^{(qk)_{+}} \!\! d(qk)
\nonumber \\
\times & \! \int\limits_{(lk)_{-}}^{(lk)_{+}} \!\! d(lk) \, A(k^2,
qk, lk) \;.
\end{align}
The dimensionless quantity $n_g(k^2)$ in \eqref{mult_cl_repres} is
the multiplicity of hadrons in the gluon jet whose virtuality is
$k^2$, while the normalization $H$ is given by
\begin{equation}\label{norm_W}
H = \frac{1}{6 \pi^2} \, m_t^4 (1- r)^2(1 + 2r) \;.
\end{equation}

The analytic expressions for the functions $D_1(k^2, qk)$,
$D_2(k^2, qk)$ in \eqref{A} are rather complicated to be shown
here.%
\footnote{They depend also on the masses $m_c$ and $m_W$.}
That is why we present only the final results of our QCD
calculations based on the formulae of this Appendix (see
Eqs.~\eqref{delta_N_Ql}, \eqref{delta_E_Ql} in the main text).

\setcounter{equation}{0}
\renewcommand{\theequation}{B.\arabic{equation}}

\section*{Appendix B}

 In this Appendix we will demonstrate that the account
of the $W$ boson width results in only small corrections to the
hadronic multiplicities.

The denominator of the $W$ boson propagator (see the diagram in
Fig.~\ref{fig:A1}) is
\begin{equation}\label{B-W}
q^2 - m_W^2 + i m_W \Gamma_W \;,
\end{equation}
where $m_W$ is the mass of the $W$ boson, $\Gamma_W$ its full
width. The mean multiplicity is given by the formula
\begin{equation}\label{mean_mult}
\langle n_h \rangle =  \frac{1}{N} \int\limits_0^{(m_t - m_b)^2}
\!\!\! d q^2 \, \frac{m_W \Gamma_W }{(q^2 - m_W^2)^2 + m_W^2
\Gamma_W^2} \, F(q^2) \, n_h(q^2)  \;,
\end{equation}
where $n_h$ ($n_h = n_W \mathrm{\ or \ } n_{tb}$, see the main
text) depends on the $W$ boson virtuality $q^2$. Here
\begin{equation}\label{factor}
F(q^2) = (m_t^2 - q^2)^2 (m_t^2 + 2q^2) \;,
\end{equation}
and the normalization $N$ is
\begin{equation}\label{norm}
N = \int\limits_0^{\infty} \! d q^2 \, \frac{m_W \Gamma_W }{(q^2 -
m_W^2)^2 + m_W^2 \Gamma_W^2} \, F(q^2) \;.
\end{equation}
In both $\langle n_h \rangle$~\eqref{mean_mult} and
$N$~\eqref{factor} the factor $m_W \Gamma_W$ is introduced, while
common constants are omitted.

In the zero width limit,
\begin{equation}\label{zero_width}
\frac{m_W \Gamma_W }{(q^2 - m_W^2)^2 + m_W^2 \Gamma_W^2} \
\xrightarrow{\Gamma_W \rightarrow 0} \ \pi \, \delta (q^2 - m_W^2)
\;,
\end{equation}
we do obtain that the mean multiplicity is equal to $n_h(m_W^2)$.

The numerical calculations with the use of formulae
\eqref{mean_mult}, \eqref{factor} result in the following values:
\begin{align}
\langle n_W \rangle &= 19.04 \;, \label{n_W_corr} \\
\langle n_{tb} \rangle &= 16.37 \;. \label{n_tb_corr}
\end{align}
Thus, the account of non-zero width of the $W$ boson slightly
changes the average value of the multiplicities. Namely,
$n_W$~\eqref{n_W_num} has gone down by 0.3, while
$n_{tb}$~\eqref{n_tb_num} has gone up by 0.23, but their sum
remains almost unchanged.


\end{document}